\documentclass[a4paper,10pt]{article}
\usepackage{hyperref}
\usepackage{url,graphicx}
\usepackage{amsthm,amsmath,amssymb,amsfonts,amscd,amsbsy,amstext}
\usepackage{showexpl}
\usepackage{mathrsfs,xr}
\usepackage[all,arc,curve,color,frame]{xy}
\usepackage{graphics,fullpage}
\usepackage{latexsym,delarray}
\xyoption{all}

\title{Spectral action for a one-parameter family of Dirac-type operators on $SU(2)$ and its inflation model}

\author{Alan Lai, Kevin Teh
\\
California Institute of Technology
}
\begin{document}
\maketitle


\newcommand{\cch}
{\widetilde{\mathtt{Ch}}} 
\newcommand{\Dd}{\mathcal{D}}
\newcommand{\sDd}{\slash\hspace{-0.26cm}\mathcal{D}}
\newcommand{\dDd}{\Dd_I}
\newcommand{\CCc}{\overline{\Cc}^{\Ggamma}}
\newcommand{\TR}{\operatorname{Tr}}
\newcommand{\Tr}{\operatorname{tr}}
\newcommand{\CCHERN}{\operatorname{ch}}
\newcommand{\KKEI}{\operatorname{K}}
\newcommand{\g}{\mathfrak{g}} 
\newcommand{\so}{\mathfrak{so}}
\newcommand{\spin}{\mathfrak{spin}}
\newcommand{\naturalnumber}{\mathbb{N}}
\newcommand{\realnumber}{\mathbb{R}}
\newcommand{\complexnumber}{\mathbb{C}}
\newcommand{\CAS}{\operatorname{Cas}}
\newcommand{\MAP}{\operatorname{Map}}
\newcommand{\GAU}{\operatorname{Gau}}
\newcommand{\MATRIX}{\operatorname{Mat_\mathbb{C}(N)}}
\newcommand{\END}{\operatorname{End}}
\newcommand{\Uu}{\mathcal{U}}
\newcommand{\Ww}{\mathcal{W}}
\newcommand{\Aa}{\mathcal{A}}
\newcommand{\Bb}{\mathcal{B}}
\newcommand{\Hh}{\mathcal{H}}
\newcommand{\Ss}{\mathcal{S}}
\newcommand{\Ff}{\mathfrak{F}}
\newcommand{\Nn}{\mathcal{N}}
\newcommand{\Pp}{\mathcal{P}}
\newcommand{\Tt}{\mathcal{T}}
\newcommand{\Ll}{\mathcal{L}_{\Nn}}
\newcommand{\Kk}{\mathcal{K}_{\Nn}}
\newcommand{\Mm}{\mathfrak{m}}
\newcommand{\SUPP}{\operatorname{supp}}
\newcommand{\HE}{\operatorname{HE}}
\newcommand{\KY}{\operatorname{K}}
\newcommand{\ED}{\mathfrak{e}}
\newcommand{\VE}{\mathfrak{v}}
\newcommand{\SO}{\operatorname{s}}
\newcommand{\RA}{\operatorname{r}}
\newcommand{\ID}{\operatorname{id}}
\newcommand{\HOM}{\operatorname{Hom}}
\newcommand{\HOL}{\operatorname{Hol}}
\newcommand{\ad}{\operatorname{ad}}
\newcommand{\Zz}{\mathbb{Z}}
\newcommand{\Cc}{\mathfrak{C}}
\newcommand{\CL}{\mathbb{C} \operatorname{ l}}
\newcommand{\Hhg}{\mathcal{G}}
\newcommand{\Gg}{\mathcal{G}}
\newcommand{\Ggamma}{\mathbf{\Gamma}}
\newcommand{\KKi}{\mathbf{\chi}}
\newcommand{\Ch}{\mathtt{Ch}_{^{_{\operatorname{JLO}}}}}
\newcommand{\ch}{\mathtt{ch}}
\newcommand{\hCh}{\mathtt{Ch}_{^{_{\operatorname{JLO}}}}}
\newcommand{\Ilim}{\varinjlim}
\newcommand{\Plim}{\varprojlim}
\newcommand{\LG}{\mathcal{LG}}
\newcommand{\ZO}{\diamond}
\newcommand{\KER}{\operatorname{ker}}
\newcommand{\DIFF}{\operatorname{diff}}
\newcommand{\ENDO}{\operatorname{End}}
\newcommand{\DOM}{\operatorname{Dom}}
\newcommand{\KCYCLE}{(\Bb,\Nn,\Dd)}
\newcommand{\FREDMOD}{(\rho,\Nn,F)}
\newcommand{\IND}{\mathtt{Ind}_\tau}
\newtheorem{definition}{Definition}[section]
\newtheorem{example}{Example}[section]
\newtheorem{theorem}{Theorem}[section]
\newtheorem{lemma}[theorem]{Lemma}
\newtheorem{corollary}[theorem]{Corollary}
\newtheorem{proposition}[theorem]{Proposition}
\newtheorem{remark}[definition]{Remark}

\setcounter{section}{0}

\begin{abstract}
We analyze the Dirac Laplacian of a one-parameter family of Dirac operators on a compact Lie group, which includes the Levi-Civita, cubic, and trivial Dirac operators. More specifically, we describe the Dirac Laplacian action on any Clifford module in terms of the action of the Lie algebra's Casimir element on finite-dimensional irreducible representations of the Lie group. Using this description of the Dirac Laplacian,  we explicitly compute spectrum for the one-parameter family of Dirac Laplacians on $SU(2)$, and then using the Poisson summation formula, the full asymptotic expansion of the spectral action. The technique used to explicitly compute the spectrum applies more generally to any Lie group where one can concretely describe the weights and corresponding irreducible representations, as well as decompose tensor products of an irreducible representation with the Weyl representation into irreducible components. Using the full asymptotic expansion of the spectral action, we generate the inflation potential and slow-roll parameters for the corresponding pure gravity inflationary theory.
\end{abstract}

\section{One-parameter family of Dirac-type operators $\Dd_t$}
In this section, we follow Agricola \cite{agricola} to define a family of Dirac operators, which interpolates
the Levi-Civita, cubic, and trivial Dirac operators.

Let the Lie algebra $\g$ of $G$ be identified as the space of left invariant tangent vector fields on $G$, then
one obtains a family of connections 
\[
\nabla _X^t := \nabla _X^0  + t [X,\cdot] ,
\]
where $\nabla^0$ is the trivial connection with respect to the left trivialization, and
$\nabla^1$ is the trivial connection with respect to the right trivialization.

Let $\langle \cdot, \cdot\rangle$ denote a positive definite invariant metric on $\g$. One
checks that $\nabla^t$ is a metric $\so(\g)$ connection, and the torsion, $T(X,Y)=(2t-1)[X,Y]$, vanishes when $t=1/2$. Hence one can think of the Levi-Civita connection, $\nabla^{1/2}$, as the middle connection in the linear interpolation between the
left trivial connection and the right trivial connection.

The $\so(\g)$ connection $\nabla^t$ lifts to a metric $\spin(\g)$ connection $\widehat{\nabla^t}$ \cite{par} given by the formula
\[
\widehat{\nabla^t}_X  = \nabla_X^0  + t \frac{1}{4}\sum_{k,l} \langle  X,[X_k, X_l]\rangle X_k X_l .
\]

Let $\CL(\g)$ denote the Clifford algebra generated by $\g$ with the relation $XY+YX=-2 \langle X,Y\rangle$
for $X,Y \in \g$. 
Let $\{X_i\}$ denote the set of orthonormal basis of $\g$ with respect to the metric $\langle \cdot, \cdot \rangle$.
Let us define  the Dirac operator in $\CL(\g)\otimes \Uu(g)$  given by $\widehat{\nabla^t}$ to be

 \begin{eqnarray}
 \label{defn:dirac}
 \Dd_t:= \sum_i X_i \otimes X_i
+ t  H \in \CL(\g) \otimes \Uu(\g) ,
\end{eqnarray}
where 
\[H:= \frac{1}{4}\sum _{j,k,l} X_j X_k X_l \otimes \langle X_j,[X_k, X_l ]\rangle \in \CL(\g) \otimes \Uu(\g) .\]

Since the connection is metric, and it is torsion free at $t=1/2$, $\Dd_{1/2}$ is the Levi-Civita Dirac operator.
$\Dd_0$ is the trivial Dirac operator defined by the trivial connection with respect to the left trivialization, it is the Dirac operator used as a basic construction block of the spectral triple of Aastrup-Grimstrup-Nest in loop quantum gravity to model tetrads \cite{AGN,lai}. The Dirac operator, $\Dd_{1/3}$, with parameter $t=1/3$  is the cubic Dirac operator of Kostant \cite{cubic}, whose square, we will see in a moment, consists of only the degree two term and the degree zero term.
Therefore, the family of Dirac operators $\Dd_t$ interpolates the most important Dirac operators one considers on a Lie group.

\section{Spectrum of $\Dd_t^2$}

We analyze the operator $\Dd_t^2$ and find that it can be written as a combination of the Casimir
operator $\CAS$. As the action of $\CAS$ on (finite dimensional) irreducible components of a Lie algebra representation is known, we reduce the study of $\Dd_t^2$ to the study of $\CAS$ action on irreducible components, and give an expression of the spectrum of $\Dd_t^2$ in terms of Lie algebra representations.

The calculation for $\Dd_t^2$ is done in \cite{agricola}. To be self-contained, we
 include the calculation for our simplified case here.

We calculate
\begin{eqnarray*}
\Dd_t^2 &=& \left(   \sum_i X_i \otimes X_i + tH  \right)^2\\
&=& \left( \sum_i X_i \otimes X_i \right)^2 + t 
\left(\sum_i X_i \otimes X_i \right) H + t H \left( \sum_i X_i \otimes X_i \right) + (tH)^2.
\end{eqnarray*}

The first term is 
\begin{eqnarray*}
\left( \sum_i X_i \otimes X_i  \right) ^2&=& \sum _i X_i X_i \otimes X_i X_i  +
\sum_{i\neq j} X_i X_j \otimes X_i X_j\\
&=& \sum_i -1 \otimes X_i^2 + \frac{1}{2} 
\sum_{i\neq j} X_i X_j \otimes X_i X_j -
\frac{1}{2} \sum_{i\neq j} X_i X_j \otimes X_j X_i\\
&=& 1 \otimes \CAS + \frac{1}{2} \sum_{ij} X_i X_j \otimes [X_i, X_j].
\end{eqnarray*}

For the cross term, first observe that
for $i \neq j, k , l$, \[X_i  X_j X_k X_l + X_j X_k X_l X_i=0\] by the Clifford relation.
Therefore, $i$ must equal one of $j,k,l$.

For $i = j$, 
\begin{eqnarray*}
\sum _{i, j,k,l}
\left( X_i  X_j X_k X_l + X_j X_k X_l X_i \right) \otimes \langle [X_j,X_k], X_l \rangle X_j
&=&  
\sum _{j,k,l} 
-2 X_k X_l  \otimes \langle [X_j,X_k], X_l \rangle X_j \\
&=& \sum_{k,l} -2 X_k X_l \otimes [X_k, X_l] .
\end{eqnarray*}

The same expression is obtained for $i=k$ and $i=l$, thus summing over the three possibilities, one obtains
\begin{eqnarray*}
 t 
\left(\sum_i X_i \otimes X_i \right) H + t H \left( \sum_i X_i \otimes X_i \right) &=&
t \frac{1}{4} 
\sum _{i, j,k,l}
\left( X_i  X_j X_k X_l + X_j X_k X_l X_i \right) \otimes \langle [X_j,X_k], X_l \rangle X_i \\
&=& - 3t \frac{1}{2} \sum_{k,l} X_k X_l \otimes [ X_k, X_l] .
\end{eqnarray*}

For the scalar term 
\begin{eqnarray}
\label{eqn:scalerterm1}
(tH)^2 = t^2 \frac{1}{16} \sum_{b,c,d,j,k,l } X_b X_c X_d X_j X_k X_l \otimes 
\langle [X_b,X_c], X_d \rangle \langle [X_j,X_k], X_l \rangle .
\end{eqnarray}

If $b\neq j,k, $ or $l$, then the above expression~\eqref{eqn:scalerterm1} is zero by the Clifford relation.

For $b=j, k, $ or $l$, 
\begin{eqnarray}
\label{eqn:scalerterm2}
\sum_{b,c,d,j,k,l } X_b X_c X_d X_j X_k X_l \otimes 
\langle [X_b,X_c], X_d \rangle \langle [X_j,X_k], X_l \rangle &=&
-3\sum_{c,d,k,l } X_c X_d  X_k X_l \otimes 
\langle [X_c, X_d] ,[X_k, X_l] \rangle .
\end{eqnarray}

If $c\neq k$ or $l$, then the above expression~\eqref{eqn:scalerterm2} is zero by the Clifford relation and
Jacobi identity. When $c=k,$ or $l$, $b$ has to equal $l,$ or $k$.
Thus
\begin{eqnarray*}
-3\sum_{c,d,k,l } X_c X_d  X_k X_l \otimes 
\langle [X_c, X_d] ,[X_k, X_l] \rangle &=& 6 \sum_{k,l} \langle [X_k,X_l],[X_k,X_l] \rangle \\
&=& 6 \sum_{k,l} \langle X_k, - [X_l, [X_l, X_k]] \rangle \\
&=& 6 \sum_k \langle X_k, \ad_{\CAS} X_k \rangle \\
&=& 6 \Tr (\ad_{\CAS}) 
.
\end{eqnarray*}

Together with the generalized strange formula of Freudenthal and de Vries
by Kostant \cite{kostant}
\[
\frac{1}{24}\Tr(\ad_{\CAS}) = \lvert \rho \rvert ^2,
\]
the scalar term~\eqref{eqn:scalerterm1} is found to be
\begin{eqnarray*}
(tH)^2
&=& t^2 \frac{3}{8} \Tr ( \ad_{\CAS})\\
&=& 9 t^2 \lvert \rho \rvert ^2.
\end{eqnarray*}


Therefore, the total expression of $\Dd_t^2$ can be written as the following.
\begin{proposition}[\cite{agricola}]
Let $\{X_i\}$ be the set of orthonormal basis of $\g$, $\Dd_t$ to be defined as in Equation~\eqref{defn:dirac}.
Then
\begin{eqnarray*}
\Dd_t^2 &=&  1\otimes \CAS + (1-3t) \frac{1}{2} \sum_{k,l}
X_k X_l \otimes [X_k, X_l] 
+ 9t^2 \lvert \rho \rvert ^2 .
\end{eqnarray*}
\end{proposition}
Notice that for $t=1/3$, the degree one term vanishes. $\Dd_{1/3}$ is called the cubic Dirac operator of Kostant \cite{cubic}.

The degree two term is written as a Casimir, however the degree one term is still rather complicated.
As it turns out, we can further simplify the expression and write the degree one term in terms of Casimir.


\begin{theorem}[\cite{kostant}]
The map $\pi: \g \to \CL(\g)$ given in the orthonormal basis $X_k\in \g$ by
\[
\pi (X_i ) := \frac{1}{4} \langle X_i,[X_k,X_l] \rangle X_k X_l
\]
is a Lie algebra homomorphism.
\end{theorem}
We extend $\pi$ to $\pi : \Uu(\g) \to \CL(\g)$.

Now the degree one term in $\Dd_t^2$ can be written as
\begin{eqnarray*}
\frac{1}{2} \sum_{k,l}
X_k X_l \otimes [X_k, X_l] 
&=&  \sum_i  \frac{1}{2} \sum_{k,l} \langle [X_k, X_l], X_i \rangle  X_k X_l \otimes X_i \\
&=&  2 \sum_i \pi (X_i) \otimes X_i .
\end{eqnarray*}

Let $\Delta:\Uu(\g) \to \Uu(\g \oplus \g)= \Uu(\g) \otimes \Uu(\g)$ denote the co-multiplication given by
the diagonal embedding \[\Delta (X_i) := X_i \otimes 1 + 1\otimes X_i \in \Uu(\g) \otimes \Uu(\g).\]
Then
\begin{eqnarray*}
\Delta \CAS &=& - \sum_i  \left( X_i \otimes 1 + 1 \otimes X_i \right) ^2\\
&=& 1 \otimes \CAS + \CAS \otimes 1 -2 \sum_i X_i \otimes X_i .
\end{eqnarray*}
Thus, 
\begin{eqnarray*}
-\frac{1}{2} \sum_{k,l}
X_k X_l \otimes [X_k, X_l] &=&  (\pi \otimes 1 )
\left(  \Delta \CAS - 1\otimes \CAS - \CAS\otimes 1\right).
\end{eqnarray*}

We obtain
\begin{theorem}
For $\Dd_t$ defined as in Equation~\eqref{defn:dirac},
$\Dd_t^2$ can be written as
\[\Dd_t^2 = (\pi \otimes 1) T_t\]
where
\begin{eqnarray*}
\Uu(\g)\otimes \Uu(\g) \ni T_t:= 1\otimes \CAS + (3t-1)  ( \Delta \CAS - 1\otimes \CAS - \CAS \otimes 1) +
9t^2 \lvert \rho \rvert^2 .
\end{eqnarray*}
\end{theorem}
Hence, one obtains the spectrum of $\Dd_t^2$ as the spectrum of $T_t$, which is written as a combination 
of $1\otimes \CAS$, $\CAS\otimes 1$, and $\Delta \CAS$.

\begin{lemma}[\cite{knapp}]
\label{lem:cas}
Let $V_\lambda$ be an irreducible representation of $\g$ with highest weight $\lambda$. 
Then $\CAS$ acts as a scalar on $V_\lambda$, and the scalar is given by
\[
\lvert \lambda + \rho \rvert ^2 - \lvert \rho \rvert ^2 = \langle \lambda + 2\rho, \lambda \rangle .
\]
\end{lemma}

Suppose that $\Ss$ is a Clifford $\CL(\g)$-module so that
$\CL(\g)\otimes \Uu(\g)$ acts on $\Ss \otimes L^2(G)$, where
$\CL(\g)$ acts on $\Ss$ via the Clifford action and $\Uu(\g)$ acts on $L^2(G)$ as
left invariant differentiation.

By Peter-Weyl, $L^2(G)$ decomposes as the norm closure of 
\[
\bigoplus _{\lambda \in \widehat{G}} V_\lambda \otimes V_\lambda ^* ,
\]
where $\lambda$ ranges over the irreducible representations $\widehat{G}$ of $G$.

Since $\Uu(\g)$ acts as left invariant differential operators on $L^2(G)$, it acts as the identity
on the dual components $V_\lambda^*$.

\begin{theorem}[\cite{kostant}]
Let $\Ss$ be any $\CL(\g)$-module. Then the $\g$-representation on $\Ss$ defined by 
composition with $\pi$
is a direct sum of $\rho$-representations, where $\rho$ is the half sum of all positive roots, the Weyl vector.
\end{theorem}

The theorem implies that as an $\Uu(\g)\otimes \Uu(\g)$ representation, $\Ss\otimes L^2(G)$ decomposes
as 
\[
\Ss\otimes L^2(G)= \bigoplus _\rho  \bigoplus _{\lambda \in \widehat{G} }
V_\rho \otimes V_\lambda \otimes V_\lambda^* .
\]
The sum $\bigoplus_\rho$ and the dual components $V_\lambda^*$ only change the
multiplicity of the action. We will for the moment ignore them and only look at the
$V_\rho \otimes V_\lambda$ part.

We can read off how $\CAS\otimes 1 $ and $1\otimes \CAS$ act on $V_\rho \otimes V_\lambda$ already
by directly applying Lemma~\ref{lem:cas}.
However to know how $T_t$, hence $\Dd_t^2$, acts on $V_\rho \otimes V_\lambda$, we need to know
the action of $\Delta \CAS$, which can be obtained if one knows the direct sum decomposition of
$V_\rho \otimes V_\lambda$ into irreducible components, the so-called Clebsch-Gordon decomposition.
In effect, we are reducing the study of the spectrum $\Dd_t^2$ to the Clebsch-Gordon decomposition of
$V_\rho \otimes V_\lambda$.

Suppose that $V_\rho\otimes V_\lambda$ decomposes as
\begin{eqnarray}
\label{eqn:cg}
V_\rho \otimes V_\lambda = \bigoplus _{\gamma} V_{\lambda\pm \gamma}
\end{eqnarray}
for some weights $\gamma$.

We examine on it the action of 
\[
T_t= 1\otimes \CAS + (3t-1)  ( \Delta \CAS - 1\otimes \CAS - \CAS \otimes 1) +
9t^2 \lvert \rho \rvert^2 .
\]

One has
\begin{eqnarray*}
-(3t-1)\CAS \otimes  1 + 9t^2 \lvert \rho \rvert^2 \Big |  _{V_\rho \otimes V_\lambda }
&=& -(3t-1) \langle \rho + 2\rho , \rho \rangle + 9t^2 \lvert \rho \rvert ^2 \\
&=& (3t-1)(3t-2) \lvert \rho \rvert ^2 + \lvert \rho \rvert ^2.
\end{eqnarray*}
and
\begin{eqnarray*}
(3t-1)\Delta \CAS - 1\otimes \CAS \Big |  _{V_\rho \otimes V_\lambda }
&=& (3t-1)\Big(
\langle \lambda \pm \gamma  + 2\rho , \lambda \pm \gamma \rangle  - 
\langle \lambda +2\rho, \lambda \rangle \Big) \\
&=& 2 (3t-1)\langle \lambda +  \rho , \pm \gamma\rangle + (3t-1)
\lvert \gamma\rvert^2\\
&=& 2 \langle \lambda +\rho, \pm (3t-1)\gamma\rangle + \lvert (3t-1)\gamma\rvert^2 - (3t-1)(3t-2)
\lvert \gamma \rvert ^2 .
\end{eqnarray*}

Thus, the action of $T_t$ on $V_\rho \otimes V_\lambda$ is given by the following.
\begin{proposition}
Let $\gamma$ denote the weights in the Clebsch-Gordon decomposition of $V_\rho \otimes V_\lambda$ as in Equation~\eqref{eqn:cg}. Then
\begin{eqnarray}
\label{eqn:eigenvalue}
T_t
\Big |  _{V_\rho \otimes V_\lambda }
=
\left \lvert
\lambda + \rho \pm \gamma (3t-1) 
\right \rvert ^2
+
(3t-1)(3t-2) \left(
\lvert \rho \rvert ^2 - \lvert \gamma \rvert ^2
\right).
\end{eqnarray}
\end{proposition}

\section{Spectral action of $\Dd_t$ on $SU(2)$}

From the previous section, we reduce the problem of finding the spectrum of $\Dd_t^2$ to 
finding the Clebsch-Gordon decomposition of $V_\rho \otimes V_\lambda$.
This section will focus on the specific case of $SU(2)$.

In the case of $SU(2)$,
$\lambda = m \in \naturalnumber$, $\rho =1$, $\gamma =1 $, and
\[
V_1 \otimes V_m  =V_{m+1} \oplus V_{m-1},
\]
where $V_{-1}$ is understood to be trivial.
The  multiplicity is 
$(m+1 \pm 1 )(m+1)$.
The Clifford module $\Ss$ equals just a single copy of $V_1$.

\begin{remark}
In the case of $SU(3)$, $\lambda = (p,q) \in \naturalnumber \times \naturalnumber$, $\rho=(1,1)$,
$\gamma = (a,b) = (0,0), (1,1), (2,-1), \mbox{ or } (-1,2)$; and its multiplicity is
$\frac{1}{4} (p+1 \pm a)(p+1) (q+1 \pm b)(q+1)(p+q +2 \pm (a+b)) (p+q+2)$. And $\Ss = V_{(1,1)}$.
\end{remark}

Plugging $\lambda =m$, $\rho =1 $, and $\gamma =1 $ into Equation~\eqref{eqn:eigenvalue},
one obtains 
the action of $T_t$ is
$m+3t$ on $V_{m+1}$ with multiplicity $(m+2)(m+1)$ for $m \geq 0$;
and $m+2-3t$ on $V_{m-1}$ with multiplicity $m(m+1)$ for $m \geq 1$, which can alternatively be written
as $-n-3t$ with multiplicity $(n+2)(n+1)$ for $n\leq -1$ by the change of indices $m+2=-n$.
Hence, the spectrum of $\Dd_t^2$ is given by $(n+3t)^2$ for $n \in \Zz$ with multiplicity $(n+2)(n+1)$.
And
the spectrum action $\TR f \left( \frac{\Dd_t^2}{\Lambda^2} \right) $ is 
\begin{eqnarray}
\label{eqn:spectralaction1}
\TR f \left( \frac{\Dd_t^2}{\Lambda^2} \right) &=& \sum _{n\in\Zz} (n+2)(n+1) f 
\left( \frac{(n+3t)^2}{\Lambda^2} \right).
\end{eqnarray}

Now we follow the analysis of Chamseddine and Connes \cite{uncanny}.
Let $g(u)=(u+2)(u+1) f\left( \frac{(u+3t)^2}{\Lambda^2} \right)$,
 its Fourier transform, denoted by $\widehat{g}(x)$, is
\begin{eqnarray*}
\widehat{g}(x)&=&\int _{\realnumber} (u+2)(u+1) f\left( \frac{(u+3t)^2}{\Lambda^2} \right) e^{-2\pi i x u} du\\
&=& \int _{\realnumber} (\Lambda y - (3t-1) ) (\Lambda y - (3t-2) ) f (y^2) e^{-2\pi i x (\Lambda y -3t) } \Lambda dy\\
&=&
\Lambda ^3  e^{-2\pi i x (-3t) } \int_{\realnumber} y^2  f(y^2) e^{-2\pi i x \Lambda y} dy \\ &&
-
\Lambda ^2 e^{-2\pi i x (-3t)} \int_{\realnumber}  3(2t-1) y f(y^2) e^{-2\pi i x \Lambda y} dy \\
&&
+
\Lambda \int_{\realnumber} (3t-1)(3t-2) f(y^2) e^{-2\pi i x \Lambda y} dy .
\end{eqnarray*}

Now let $\widehat{f}^{(m)}$ denote the Fourier transform of $y^mf(y^2)$, and by the Poisson summation formula
$\displaystyle 
\sum _{\Zz} g(n) = \sum_{\Zz} \widehat{g}(x)$.
The spectral action $\TR f \left( \frac{\Dd_t^2}{\Lambda^2} \right) $ \eqref{eqn:spectralaction1} then becomes
\begin{eqnarray*}
\TR f \left( \frac{\Dd_t^2}{\Lambda^2} \right)  &=& \sum_{\Zz} \widehat{g}(n)\\
&=& \Lambda^3 \sum_{\Zz}  e^{-2\pi i n (-3t) }\widehat{f}^{(2)}(\Lambda n) \\&& -
\Lambda^2  \sum_{\Zz}
3(2t-1)  e^{-2\pi i n (-3t) }\widehat{f}^{(1)}(\Lambda n) \\&&
+\Lambda\sum_{\Zz}
 (3t-1)(3t-2) e^{-2\pi i n (-3t) } \widehat{f}(\Lambda n). 
\end{eqnarray*}

By taking $f$ to be 
a Schwartz function,  $\widehat{f}^{(m)}$ has rapid decay, thus for all $k$
\[
\left\lvert \widehat{f}^{(m)} (\Lambda n)  \right\rvert < C_k(\Lambda n )^{-k}
\]
and
\[
\left\lvert \sum_{n\neq 0 } \widehat{f}^{(m)} (\Lambda n)  \right\rvert < C'_k \Lambda ^{-k}.
\]
As a result,
\[
\sum _{n\neq 0} \widehat{g}(n) \in O( \Lambda ^{-\infty}).
\]
Finally, we obtain
\begin{theorem}
The spectral action of $\Dd_t$ for $SU(2)$ is
\begin{eqnarray*}
\TR f \left( \frac{\Dd_t^2}{\Lambda^2} \right) &=&
\widehat{g}(0) + O(\Lambda ^{-\infty})\\
&=& \Lambda ^3 \widehat{f}^{2}(0) - \Lambda^2  \widehat{f}^{1}(0) 3(2t-1) +
\Lambda \widehat{f}(0) (3t-1)(3t-2)+O(\Lambda ^{-\infty}).\\
&=& \Lambda^3 \int _{\realnumber} y^2 f(y^2) dy +
\Lambda  (3t-1)(3t-2)\int_{\realnumber} f(y^2) dy  + O(\Lambda ^{-\infty}).
\end{eqnarray*}
\end{theorem}

Observe that for the cubic Dirac operator ($t=1/3$), the coefficient of the  $\Lambda$ term is zero.
As $SU(2)$ is three dimensional, this coefficient is the second Seeley-De Witt coefficient, which
depicts the generalized Einstein-Hilbert action. Therefore, the cubic Dirac operator gives an identically vanishing Einstein-Hilbert action.

\section{Slow roll potential and cosmic topology}
The above nonperturbative computation of the spectral action for the one-parameter family of Dirac operators gives an expression which naturally lends itself to the consideration of the inflation model studied in the recent papers \cite{MPT1} \cite{MPT2} \cite{CMT}.  

In order to generate an inflation scenario from the non-perturbative spectral action, one computes the non-perturbative spectral action of the Friedmann metric on the product of $S^3$ with radius $a$ and a compactified time dimension in the form of a circle with radius $\beta$.  The one-parameter family of Dirac operators above makes available for study a one-parameter family of Dirac operators on the product space-time. This inflation is a toy model insofar that the geometry considered does not include the matter sector. It is the same model as that of \cite{MPT1}. A more sophisticated analysis would require consideration of an almost commutative geometry, including an analysis of the resulting twisted Dirac operators, for instance as in \cite{CMT}. 

In the model that we consider, the family of Dirac operators on the product space-time is given by

\[ 
\Dd = \left( \begin{array}{cc}
0 & \Dd_t \otimes 1 + i \otimes \Dd_s \\
\Dd_t \otimes 1 - i \otimes \Dd_s & 0 \\
 \end{array} \right)
 \] 
 where $\Dd_s$ is the  Dirac operator on $S^1$ acting on the set of $L^2$-sections 
 $\{f: \realnumber \to \complexnumber \lvert f(x+1)= e^{2\pi i s } f(x) \}$  of the
 complex line bundle $L_s$.
 The spectrum of $\Dd_s$ is 
  $\{  \mathbb{Z} +s\}$ with each point having multiplicity $1$.  The square is then given by

\[ 
\Dd _{(t,s)}^2 = \left( \begin{array}{cc}
 \Dd_t ^2 \otimes 1 + 1 \otimes \Dd_s ^2 & 0 \\
0 & \Dd_t ^2 \otimes 1 + 1 \otimes \Dd_s ^2 \\
 \end{array} \right).
 \]
  
 One may compute the spectral action in its non-perturbative form for the above Dirac operator using the Poisson summation formula in two dimensions. First observe that the eigenvalues of $\Dd_{(t,s)} ^2 / \Lambda^2$ are
 \[
 (n+3t)^2/ (\Lambda a)^2 + (m+s)^2/ (\Lambda \beta)^2
 \] 
 and for $n, m \in \mathbb{Z}$, the eigenvalue appears with multiplicity $2(n+1)(n+2)$. 
 Taking $u = n+3t $ and $v= m+s$, we define 
 \[
 g(u,v) = 2\left(u-(3t-1)\right) \left(u-(3t-2)\right)h\left(u^2/(\Lambda a)^2 + v^2/(\Lambda \beta)^2\right)
 \]
 so that the spectral action becomes
 
 \[
 \TR h\left(\Dd_{(t,s)}^2/ \Lambda^2\right) = \sum_{(u,v) \in \mathbb{Z}^2} g(u - 3t, v - s).
 \]
 Applying the Poisson summation formula in two variables and some simple estimates, the spectral action, up to a term of order $O(\Lambda^{-k})$ for any $k$, is equal to $\widehat{g}(0,0)$. Computing $\widehat{g}(0,0)$ reveals
 \begin{eqnarray*}
  \widehat{g}(0,0) &=&
  2\Lambda \beta (\Lambda a)^3 \int_{\mathbb{R}^2} x^2 h(x^2+y^2)dx dy 
  + 2(3t-1)(3t-2)\Lambda \beta (\Lambda a) \int_{\mathbb{R}^2}h(x^2+y^2)dx dy 
 \end{eqnarray*}
 By converting to polar coordinates, one obtains the following:
 \begin{theorem}
 The spectral action of $\Dd_{(t,s)}$ on the above product $S^3 \times S^1$ is equal to
 \[
  \TR h\left(\Dd_{(t,s)}^2/ \Lambda^2\right) = \pi \Lambda^4 \beta a^3 \int_0^{\infty} r h(r^2)dr + 2\pi(3t-1)(3t-2) \Lambda^2 \beta a \int_0^{\infty} h(r^2) dr +O(\Lambda^{-k}),
 \]
 for any $k \in \mathbb{Z}$.
 \end{theorem}
 Notice that the spectral action is independent of $s$, the parameter of the complex line bundle. However, the coefficient of the $\Lambda^2$ term still depends on $t$.

Next, one analyzes the change of the spectral action under the transformation $\Dd_{(t,s)}^2 \mapsto \Dd_{(t,s)}^2 + \phi^2$. Under this transformation, the spectral action gains additional terms, amounting to a slow roll potential. The analysis of each term follows that of \cite{uncanny} and we obtain the analogous result:

\begin{theorem}
 After applying the transformation $\Dd_{(t,s)}^2 \mapsto \Dd_{(t,s)}^2 + \phi^2$, and assuming that $h$ is a smooth constant function on the interval $[0,c]$ and that $\phi^2/\Lambda^2 \leq c$, the spectral action on the above product $S^3 \times S^1$ is equal to
 \begin{align*}
  \TR h\left((\Dd_{(t,s)}^2 + \phi^2)/ \Lambda^2\right) &= \pi \Lambda^4 \beta a^3 \int_0^{\infty} r h(r^2)dr + 2\pi(3t-1)(3t-2) \Lambda^2 \beta a \int_0^{\infty} h(r^2) dr \\
  & -\pi \Lambda^2 \beta a^3 \int_0^{\infty} h(r)dr + \frac{\pi}{2}\beta a h(0) \phi^2 + \frac{\pi}{2} \beta a^3 h(0) \phi^4  - 2\pi \beta a (3t-1)(3t-2)h(0) \phi^2 + O(\Lambda^{-k}), 
 \end{align*}
 
 \end{theorem}
 
 One is also interested in the more general situation, where we do not insist that $h$ be constant on an interval or make any restrictions on the values of $\phi^2/\Lambda^2$. In this direction we have the following result:  

\begin{theorem}
 After applying the transformation $\Dd_{(t,s)}^2 \mapsto \Dd_{(t,s)}^2 + \phi^2$, and assuming that $h$ is of the form
 \[
 h(x) = P(\pi x)e^{- \pi x}, 
 \]
 where $P$ is a polynomial. We have
 \begin{align*}
  \TR h\left((\Dd_{(t,s)}^2+\phi^2)/ \Lambda^2 \right) &= \pi \Lambda^4 \beta a^3 \int_0^{\infty} r h(r^2)dr + 2\pi(3t-1)(3t-2) \Lambda^2 \beta a \int_0^{\infty} h(r^2) dr \\
  & + \pi \Lambda^4 \beta a^3 \mathcal{V}(\phi^2/\Lambda^2) -2 \pi \Lambda^2 \beta a (3t-1)(3t-2)\mathcal{W}(\phi^2/\Lambda^2) + O(\Lambda^{-k}),
   \end{align*}
  for any $k$, where the functions $\mathcal{V}$ and $\mathcal{W}$ are given by
  \[
  \mathcal{V}(x) = \int_0^{\infty}r(h(r+x)-h(r))dr, \quad \mathcal{W}(x) = \int_0^{x}h(r)dr
  \] 
  
  The corresponding slow-roll potential is then
  \[
  V(x) = \pi \Lambda^4 \beta a^3 \mathcal{V}(x) - 2 \pi \Lambda^2 \beta a (3t-1)(3t-2)\mathcal{W}(x),
  \]
  where $x = \phi^2/\Lambda^2$.
 \end{theorem}

The first and second slow-roll parameters, respectively $\epsilon$ and $\eta$, can be derived from the slow-roll potential according to the following relations:

\[
A = \frac{1}{2} \left( \frac{V'(x)}{V(x)}\right)^2 \quad \mathrm{and} \quad B = \left( \frac{V''(x)}{V(x)}\right)^2 
\]

\[
\epsilon = \frac{m_{Pl}^2}{8 \pi}A \quad \mathrm{and} \quad \eta = \frac{m_{Pl}^2}{8 \pi}(B-A),
\]

where $m_{Pl}$ is the Planck mass.

The slow-roll parameters are then given by 
\begin{equation}
\epsilon = \frac{m_{Pl}^2}{16 \pi} \left( \frac{- (\Lambda a)^2 \int_x^{\infty} h(r) dr - 2(3t-1)(3t-2)h(x)}{(\Lambda a)^2 \int_0^{\infty} r(h(r+x)-h(r))dr - 2(3t-1)(3t-2)\int_0^x h(r) dr}\right)^2 
\end{equation}

\begin{align*}
\eta &= \frac{m_{Pl}^2}{8 \pi} \left( \frac{(\Lambda a)^2 h(x) - 2(3t-1)(3t-2)h'(x)}{(\Lambda a)^2 \int_0^{\infty} u(h(u+x)-h(u))du - 2(3t-1)(3t-2)\int_0^x h(u) du}\right)^2 \\
& -\frac{m_{Pl}^2}{16 \pi} \left( \frac{- (\Lambda a)^2 \int_x^{\infty} h(r) dr - 2(3t-1)(3t-2)h(x)}{(\Lambda a)^2 \int_0^{\infty} r(h(r+x)-h(r))dr - 2(3t-1)(3t-2)\int_0^x h(r) dr}\right)^2 
\end{align*}

This theory of inflation is interesting insofar as the inflation potential of the theory as well as the slow-roll parameters depend on the underlying topology, as shown in \cite{MPT2}. Here, we see that the inflation potential and slow-roll parameters are sensitive to the connection used to define the Dirac operator, in addition to the underlying topology.

\bibliographystyle{plain}


\begin{thebibliography}{99}

\bibitem{agricola} I. Agricola, {\em Connections on naturally reductive spaces, their Dirac operator and homogeneous models in string theory}.
Commun.Math.Phys. 232 (2003) 535-563.

\bibitem{uncanny} A. Chamseddine, A. Connes, {\em
The Uncanny Precision of the Spectral Action}.
	Commun.Math.Phys. 293 (2010) 867-897.

  \bibitem{AGN}
  J. Aastrup, J. M. Grimstrup and R. Nest, 
{\em On Spectral Triples in Quantum Gravity II}, J. Noncommut. Geom. 3 (2009) 47.


\bibitem{knapp} A. Knapp, {\em Lie Groups: Beyond an Introduction}.
Birkh{\"{a}}user (1996).


\bibitem{kostant} B. Kostant, {\em 
Clifford algebra analogue of the {H}opf-{K}oszul-{S}amelson theorem, 
the $\rho$-decomposition 
$\operatorname{C}(\g) = \operatorname{End}(V_{\rho})\otimes \operatorname{C}(P)$,
 and the $\g $-module 
structure of $\wedge \g $"}.
Advances in Mathematics. 2 (1997), 275-350.


\bibitem{cubic} B. Kostant, {\em A cubic Dirac operator and the emergence of Euler number multiplets of representations for equal rank subgroups}. 
 Duke Math. J. Volume 100, Number 3 (1999), 447-501.
 
 \bibitem{lai}
A. Lai, 
{\em 
The JLO Character for The Noncommutative Space of Connections of Aastrup-Grimstrup-Nest},
 arXiv:1010.5226 (2010).
 
 \bibitem{par}
 R. Parthasarathy, {\em Dirac operator and the discrete series}. Ann. of Math. 96 (1972), no. 1, 1–30.

\bibitem{deJong} J. de Jong, {\em Graphs, spectral triples and Dirac zeta functions}. P-Adic Numbers Ultrametric Anal. Appl. 1 (2009), no. 4, 286–296. 

\bibitem{MPT1} M. Marcolli, E. Pierpaoli, K. Teh, {\em The spectral action and cosmic topology}.
Comm. Math. Phys. 304 (2011), no. 1, 125–174.

\bibitem{MPT2}M. Marcolli, E. Pierpaoli, K. Teh, {\em The coupling of topology and inflation in noncommutative cosmology}. Comm. Math. Phys. 309 (2012) no. 2, 341-369

\bibitem{CMT}B. Cacic, M. Marcolli, K. Teh {\em Coupling of gravity to matter, spectral action and cosmic topology} arxiv:1106.5473, (2011) to appear in JNCG


\end{thebibliography}

\end{document}